\documentclass[twocolumn,showpacs,prl]{revtex4}
\usepackage{amsmath}
\usepackage{mathrsfs}
\usepackage{graphicx}

\begin{document}
\draft

\title{An Electronic Mach-Zehnder Quantum Eraser}%

\author{Kicheon Kang}\email{kckang@chonnam.ac.kr}
\affiliation{Department of Physics and Institute for Condensed
Matter Theory, Chonnam National University, Gwangju 500-757,
Korea}

\date{\today}

\begin{abstract}
We propose an electronic quantum eraser in which electrons are injected
into a mesoscopic conductor in the quantum Hall regime. The conductor is 
composed of a two-path interferometer, an electronic analog of the optical
Mach-Zehnder interferometer, and a quantum point contact detector 
capacitively coupled to the interferometer. 
While the interference of 
the output current at the interferometer is suppressed by
the {\em which-path} information, we show that the which-path information
is erased and the interference reappears in the cross correlation 
measurement between 
the interferometer and the detector output leads. We also investigate a 
modified setup in which the detector is replaced by a two-path interferometer.
We show that the distinguishability of the path and the visibility of 
joint
detection can be controlled in a continuous manner and satisfy a 
complementarity relation for the entangled electrons.
\end{abstract}


%%  03.65.Ta  %Foundations of quantum mechanics; measurement theory (for
%%  optical tests of quantum theory, see 42.50.Xa)
%%  03.67.-a  Quantum Information
%%  73.23.-b  % Electronic transport in mesoscopic systems

%%%%%%
%\let\veps=\varepsilon%
%%
\maketitle
Complementarity in quantum theory is well described in a double-slit
interferometer. In a two-path interferometer
with a {\em which-path} (WP) detector, the observation of interference 
and acquisition of WP information are mutually 
exclusive~\cite{feynman65,wooters79,stern90,scully91}. 
Feynman argued that any attempt to extract WP
information would disturb the motion of the injected particle and
wash out the interference pattern~\cite{feynman65}.
That is, if one can
fully determine the path, then there is no interference at all that
results from
the uncertainty of the phase of the incident particle~\cite{stern90}. 
The interference is lost even if one does not actually readout the
WP detector, as far as there is a mere possibility of carrying
out the measurement.
However, the loss of interference need not be irreversible if
the WP detector itself is a quantum system. In particular,
it has been proposed~\cite{scully91,scully82} that in some cases the loss of
interference may be simply because of the quantum correlation of the
interferometer with the WP detector. In this case the WP information can be
{\em erased} by a suitable measurement on the detector. This
`quantum eraser' has been realized in various setups by using 
entangled photons~\cite{QE-exp}.

On the other hand, mesoscopic physics is evolving into a stage where
understanding the measurement process is becoming important. Indeed,
WP detection in quantum interferometers has been achieved by
using mesoscopic conductors~\cite{buks98,sprinzak00,kalish04}.
In these experiments, a quantum point contact (QPC) was used
as a WP detector by probing the charge of a single electron
at a nearby quantum dot (QD). There have been numerous theoretical
studies on how detection of a charge suppresses the 
interference~\cite{aleiner97,levinson97,gurvitz97,hacken98,stodolsky99,buttiker00,korotkov01,pilgram02,averin05,kang05,khym06,khym06a}. 
In many theoretical works on mesoscopic
WP detection, suppression of interference due to quantum correlation
is identified with the back-action dephasing caused by the
disturbance of the electronic motion by the detector~\cite{stern90}.
However, in principle, a quantum erasure is possible also for
mesoscopic systems unless the suppression of the interference is not
accompanied by irreversible phase randomization. There was a 
theoretical proposal for a mesoscopic quantum eraser by exploiting the 
internal degree of freedom for the QD state~\cite{hacken98a}

In this Letter, we propose a new idea for a quantum eraser composed of an
electronic Mach-Zehnder interferometer (MZI)~\cite{optics84,ji03} 
and a QPC detector. 
Two different types of quantum erasers are considered. In the first setup, 
an electronic beam splitter plays the role of 
the WP detector (Fig.~1). 
The MZI and the beam splitter are capacitively
coupled in a way that the WP information is stored in the scattering
phase because of Coulomb interaction. 
An electronic analog of the beam splitter can be made by using
a QPC at high magnetic fields~\cite{buttiker90}. An electronic MZI has 
recently been realized by using quantum Hall
edge states~\cite{ji03,neder06,neder05}. Therefore it is possible to achieve 
the arrangements of Figs.~(1,2). Without the detector, 
the output currents at
lead $\alpha$ and at $\beta$ will show interference as a function of
the phase ($\varphi$) enclosed by the loop of the MZI. This phase can be
controlled either by an external magnetic field or by an electrical 
gate~\cite{ji03,neder06,neder05}. Coulomb interaction between the
electrons of the interferometer and the detector leads to a 
change in the electronic trajectory (denoted by the dashed lines), when
both electrons are transmitted. 
This modification of the trajectory gives partial or full WP information 
for the electron in the MZI and suppresses the interference. Interestingly,
the WP information can be erased by measuring the joint-detection probability
of two electrons at lead $A$ ($\in \alpha,\beta$) and 
lead $B$ ($\in\gamma,\delta$).
Furthermore, in the second setup where the detector is also made of a
MZI (Fig.~2), the path distinguishability and the visibility of the 
joint-detection probability can
be controlled by modulating the phase in the detector.
In other words, one can choose the `particle-like' or the `wave-like' 
behavior by an appropriate manipulation of the detector. 
It is important to note that our quantum eraser exploits only the
charge degree of freedom which is relatively easy to control in mesoscopic
devices.

%% From here contents %%
Let us consider a two-electron injection process (one from the
MZI and the other from the detector) in the first setup (Fig.~1).  
For convenience, two types of electron creation operators are defined, namely
$c_x^\dagger$ and $b_x^\dagger$. The operators $c_x^\dagger$ and $b_x^\dagger$
create an electron at lead $x$ and at the 
intermediate regions, respectively. 
The beam splitter, BS-$i$, is characterized by the scattering matrix
$S_i$ ($i=1,2,3$)
\begin{subequations}
 \label{eq:S-matrix}
\begin{equation}
 S_i = \left(
       \begin{array}{cc}
         r_i & t_i' \\
         t_i & r_i'
       \end{array} \right) \;,
\end{equation}
which transforms the electron operators as
\begin{eqnarray}
 ( \begin{array}{cc}
          c_{\bar{\alpha}}^\dagger & c_{\bar{\beta}}^\dagger
        \end{array} )
 =  ( \begin{array}{cc}
          b_\alpha^\dagger & b_\beta^\dagger
        \end{array} ) S_1\;, \\ 
 ( \begin{array}{cc}
          b_\alpha^\dagger & b_\beta^\dagger
        \end{array} )
 =  ( \begin{array}{cc}
          c_\alpha^\dagger & c_\beta^\dagger
        \end{array} ) S_2\;, \\ 
 ( \begin{array}{cc}
          c_{\bar{\gamma}}^\dagger & c_{\bar{\delta}}^\dagger
        \end{array} )
 =  ( \begin{array}{cc}
          c_\gamma^\dagger & c_\delta^\dagger
        \end{array} ) S_3\;.
\end{eqnarray}
\end{subequations}

Before interaction at the intermediate region, the two-electron
state can be written as
\begin{equation}
 \left|\Psi_0\right\rangle = c_{\bar{\alpha}}^\dagger 
     c_{\bar{\gamma}}^\dagger|0\rangle
      = ( r_1b_\alpha^\dagger+t_1b_\beta^\dagger )
     \otimes ( r_3c_\gamma^\dagger+t_3c_\delta^\dagger ) |0\rangle
\end{equation}
where $|0\rangle$ is the ground state without an electron injection into
the conductor.
Coulomb repulsion modifies the trajectory of the two electrons 
when both electrons are transmitted (dashed lines). 
Here it is assumed that the
Coulomb interaction affects only the trajectory of the two electrons. The
inelastic scattering is neglected. This modification of the
trajectory gives an additional phase shift $\Delta\phi$ given as  
\begin{equation}
\Delta\phi = 2\pi H\Delta A/\Phi_0 \,, 
 \label{eq:Delta-phi}
\end{equation}
where $H$ and $\Delta A$ stand for the external magnetic field and
the area enclosed by the change of the trajectory resulting from
the interaction (denoted by shaded regions of Figs.(1,2)). 
$\Phi_0$ is the flux
quantum of an electron, $\Phi_0=hc/e$. As a result, the
two-electron state upon a scattering can be written as
\begin{equation}
 \left|\Psi\right\rangle = 
   ( r_1r_3 b_\alpha^\dagger c_\gamma^\dagger 
   + r_1t_3 b_\alpha^\dagger c_\delta^\dagger 
   + t_1r_3 b_\beta^\dagger c_\gamma^\dagger 
   + t_1t_3 e^{i\Delta\phi} b_\beta^\dagger c_\delta^\dagger 
   ) |0\rangle \;,
\end{equation}
or can be simplified as 
\begin{subequations}
 \label{eq:Psi}
\begin{equation}
 \left|\Psi\right\rangle =
   ( r_1 b_\alpha^\dagger \chi_r^\dagger
   + t_1 b_\beta^\dagger \chi_t^\dagger
   ) |0\rangle \;.
\end{equation}
The operators $\chi_r^\dagger$ and $\chi_t^\dagger$ create the detector states
depending on whether the electron in the MZI is reflected or transmitted,
respectively. These operators are written as 
\begin{eqnarray}
 \chi_r^\dagger &=& r_3 c_\gamma^\dagger + t_3 c_\delta^\dagger \;, \\
 \chi_t^\dagger &=& r_3 c_\gamma^\dagger  
                 + t_3 e^{i\Delta\phi} c_\delta^\dagger \;. 
\end{eqnarray}
\end{subequations}
Eq.~(\ref{eq:Psi}) describes an entanglement between the MZI and the
detector. One can see that $\chi_r^\dagger\ne\chi_t^\dagger$ because of the
phase factor $e^{i\Delta\phi}$,
and the extent of the entanglement can be controlled by changing $H$ or
$\Delta A$.
In this way, the WP information in the MZI is stored in the detector
state.

The interference of single electrons in the MZI will be reflected in
the probability of finding an electron at lead $A$ ($\in \alpha, \beta$), 
\begin{equation}
 P_A = \langle\Psi| c_A^\dagger c_A |\Psi\rangle \;.  
\label{eq:PA}
\end{equation}
The evaluation can be done with the help of 
Eqs.~(\ref{eq:S-matrix},\ref{eq:Psi}).
The evaluation gives
\begin{eqnarray}
 P_\alpha &=& 1-P_\beta \label{eq:Pa} \\ 
   &=& R_1R_2 + T_1T_2 + 2|\nu|\sqrt{R_1T_1R_2T_2}
     \cos{(\varphi-\phi_\nu)} \nonumber\;, 
%     \cos{(\phi_{t_1}+\phi_{t_2'}-\phi_{r_1}-\phi_{r_2}-\phi_\nu)} \;, 
\end{eqnarray}
where $T_i=|t_i|^2$ and $R_i=|r_i|^2$ correspond to the transmission
and the reflection probability, respectively, for beam splitter BS-$i$.
The overlap of the detector states
$\nu\equiv \langle0|\chi_t\chi_r^\dagger|0\rangle$ is a quantitative
measure of the WP information and $\phi_\nu\equiv \arg{\nu}$. 
A small overlap indicates nearly orthogonal
detector states which distinguishes the path of an electron in the MZI.
The phase $\varphi$ enclosed by the loop of the MZI is given as 
$\varphi=\arg{(t_1)}+\arg{(t_2')}-\arg{(r_1)}
-\arg{(r_2)}$.

Eq.~(\ref{eq:Pa}) shows the relation between the interference fringe
and the WP information stored in the detector. If the two detector states
$\chi_r^\dagger|0\rangle$ and $\chi_t^\dagger|0\rangle$ are orthogonal
(that is $\nu=0$), then the electron in the MZI acquires the complete WP
information and the interference disappears. Complete WP information
can be obtained for a symmetric BS-3 ($|r_3|=|t_3|=1/\sqrt{2}$)
with $\Delta\phi=\pi$. 

Acquisition of the WP information in our setup results from 
the quantum correlation
between the two subsystems. Since the detector itself is a two-state
quantum system, the WP information can be erased by a suitable measurement on
the detector. Indeed, a joint detection of two electrons (one from a lead
of the MZI and the other from a lead of the detector) renders
a measurement
of $\Delta\phi$ impossible and therefore erases the WP
information. For simplicity, our discussion is limited to the simple case
in which the two beam splitters of the MZI are symmetric ($R_i=T_i=1/2$ for
$i=1,2$). 
The joint-detection probability $P_{AB}$ denotes
the probability of finding an electron at lead $A$ ($\in\alpha,\beta$)
and the other electron at lead $B$ ($\in\gamma,\delta$) simultaneously, 
defined as
\begin{equation}
  P_{AB} = \langle\Psi| c_A^\dagger c_A c_B^\dagger c_B |\Psi\rangle \;. 
\end{equation}
For the state $|\Psi\rangle$ defined in Eq.~(\ref{eq:Psi}), we find 
\begin{subequations}
\begin{eqnarray}
 P_{\alpha\gamma} &=& R_3 \left[ 1 + \cos{\varphi} \right]/2 \;, \\
 P_{\alpha\delta} &=& T_3 \left[ 1 + \cos{(\varphi+\Delta\phi)}
                    \right]/2 \;, \\ 
 P_{\beta\gamma} &=& R_3 \left[ 1 - \cos{\varphi} \right]/2 \;, \\
 P_{\beta\delta} &=& T_3 \left[ 1 - \cos{(\varphi+\Delta\phi)}
                    \right]/2 \;. 
\end{eqnarray}
\end{subequations}
This result shows that {\em the visibility of $P_{AB}$ is not affected by
the parameter $\nu$}. That is, the WP information encoded in the phase
shift $\Delta\phi$ is deleted by the
joint-detection measurement and the hidden coherence reappears.

Let us now consider how this effect can be verified experimentally
for a mesoscopic conductor.
At a voltage bias $eV$ for the input leads $\bar{\alpha}$ and $\bar{\gamma}$,
we can write the `entangled' many-body transport state as
\begin{equation}
 |\bar{\Psi}\rangle = \prod_{0<E<eV} \left[
   r_1 b_\alpha^\dagger(E) \chi_r^\dagger(E)
   + t_1 b_\beta^\dagger(E) \chi_t^\dagger(E)
   \right] |\bar{0}\rangle \;,   
\end{equation}
where $|\bar{0}\rangle$ stands for the ground state, a filled Fermi sea in 
all leads at energies $E<0$.
The crucial assumption made here is that the injected electrons
from the two sources interact with each other and are
transmitted as entangled pairs as illustrated in Fig.~1. 

For the state $|\bar{\Psi}\rangle$, the output current at lead $A$ 
($I_A$) is proportional to the probability of
finding an electron at this lead, 
$I_A=(e^2/h)P_AV$,
at zero temperature. The zero-frequency cross correlation, $S_{AB}$, of
the current fluctuations, $\Delta I_A$ and $\Delta I_B$, is defined 
as~\cite{buttiker90} 
\begin{equation}
 S_{AB} = \int dt\, \langle\bar{\Psi}|
   \Delta I_A(t)\Delta I_B(0)+\Delta I_B(0)\Delta I_A(t)
        |\bar{\Psi}\rangle \;,
\end{equation}
where $\Delta I_A = I_A - \langle I_A\rangle$ and 
$\Delta I_B = I_B - \langle I_B\rangle$. 
For $A\in\alpha,\beta$ and $B\in\gamma,\delta$, this cross correlator 
provides information about the two-particle interactions between the MZI
and the detector.  After some algebra, one can find the following useful
relation:
\begin{equation}
 S_{AB} = \frac{2e^2}{h} eV (P_{AB}-P_AP_B) \;.
\end{equation}
Therefore the values of the single-particle ($P_A$) and the 
joint-detection ($P_{AB}$) probabilities can be obtained by measuring 
the current and the zero-frequency cross correlation.

For the first setup shown in Fig.~1, the WP information is contained
in the detector by the entanglement of the electronic trajectories 
and is erased by measuring the coincidence count.
Next, we consider a modified setup where another beam splitter (BS-4) 
is inserted in the detector (see Fig.~2)
so that the detector itself is also a MZI. In the following, 
the upper MZI is labeled as ``MZI-$s$" and the lower MZI as ``MZI-$d$".
In this geometry, the WP information
may be erased, or marked, or partially erased by appropriate control
of the detector. The control of
the WP information is possible in a continuous manner as a function of
the phase ($\varphi_d$) enclosed by the loop of the MZI-$d$.

In addition to the scattering matrices defined in Eq.~(\ref{eq:S-matrix}),
we need another scattering matrix $S_4$ to describe the BS-4 introduced in
the same as that expressed in Eq.~(\ref{eq:S-matrix}).
The probability of finding a single electron at lead A, 
$P_A$ ($A\in\alpha,\beta$), 
is the same as that of Eq.~(\ref{eq:PA}). This is because the
quantity $\nu$ is not modified
by the unitary scattering process at the BS-4. On the other hand, the
joint-detection probability is given, for instance, as
\begin{equation}
 \label{eq:Pac}
 P_{\alpha\gamma} = |r_1r_2u_\gamma + t_1t_2'v_\gamma|^2 
                                \;,
\end{equation}
where
the coefficient $u_\gamma\equiv r_3r_4+t_3t_4'$  
($v_\gamma\equiv r_3r_4+t_3t_4'e^{i\Delta\phi}$) represents the 
amplitude of finding an electron at lead $\gamma$ under the 
condition that the electron in the MZI-$s$ passes through the 
upper (lower) path.
These coefficients can be controlled through the phase 
$\varphi_d$ enclosed by the loop of the detector given as
$\varphi_d = \phi_{t_3}+\phi_{t_4'}-\phi_{r_3}-\phi_{r_4}$. 
For example, $|u_\gamma|$ and $|v_\gamma|$
are given as
 $|u_\gamma| = |1+e^{i\varphi_d}|/2$ and  
 $|v_\gamma| = |1-e^{i\varphi_d}|$,
for the symmetric beam splitters in MZI-$d$ ($R_3=T_3=1/2, R_4=T_4=1/2$).
The other joint-detection probabilities can be
evaluated in a similar way.
The two amplitudes, $r_1r_2u_\gamma$ and $t_1t_2'v_\gamma$,
(schematically drawn in Fig.~2(b) and (c), respectively) are indistinguishable
and lead to the interference fringe for $P_{\alpha\gamma}$ with its 
visibility given by 
\begin{equation}
 \label{eq:visibility}
 {\cal V} = \frac{2\sqrt{R_1R_2T_1T_2}|u_\gamma||v_\gamma|
     }{ R_1R_2|u_\gamma|^2 + T_1T_2|v_\gamma|^2} \;.
\end{equation}
One can find that $0\leq {\cal V}\leq1$. This result shows
that the visibility of the joint-detection
probability can be controlled through the phase $\varphi_d$. In other words,
the WP information in the MZI-$s$ can be marked (${\cal V}=0$),
erased (${\cal V}=1$), or partially erased ($0<{\cal V}<1$),
by its entangled twin in the MZI-$d$.
 
The observations of the interference pattern and the acquisition of the
path information are mutually exclusive. In the following, 
we quantitatively express this interferometric duality. 
Visibility, ${\cal V}$, of Eq.~(\ref{eq:visibility}) is a measure of the
interference fringe.
On the other hand, the amount of the path information can be expressed by
the {\em distinguishability} of the two paths (Fig.~2(b,c)) defined as
\begin{equation}
 {\cal D} \equiv \frac{ \left| R_1R_2|u_\gamma|^2-T_1T_2|v_\gamma|^2 \right|
  }{ R_1R_2|u_\gamma|^2+T_1T_2|v_\gamma|^2 }
   \;.
\end{equation}
This number, which is in the range of $0\leq{\cal D}\leq1$, is a quantitative
measure of the knowledge of the paths~\cite{englert96,note}.
% Modify from here
Then, one finds the duality relation
\begin{equation}
 \label{eq:duality}
 {\cal D}^2 + {\cal V}^2 = 1 \;.
\end{equation}
The duality relation of Eq.~(\ref{eq:duality}) can be tested
experimentally. The visibility
is obtained from the interference pattern of $P_{\alpha\gamma}$ 
(Eq.~(\ref{eq:Pac})). The 
distinguishability is available in an independent measurement of 
$|u_\gamma|^2$ and $|v_\gamma|^2$. The quantities $|u_\gamma|^2$ and 
$|v_\gamma|^2$
are proportional to the output current at lead $\gamma$ in which the electron
at the BS-1 is fully reflected ($T_1=0$) or fully transmitted ($T_1=1$), 
respectively.  This independent measurement on ${\cal V}$ and ${\cal D}$ 
is expected to reveal the duality relation.

The duality relation of Eq.~(\ref{eq:duality}) is valid under the assumption
that the two injected electrons are described by an ideal pure state.
In general, the two-electron state may be
affected by the environment. In this case the duality relation is modified
as an inequality such as ${\cal D}^2+{\cal V}^2\leq1$.

Our discussion on the quantum erasure is based on the interferometer-detector 
entanglement of the electron trajectories. This kind of entanglement  
has no optical analogue since the entanglement considered here 
is formed by the Coulomb interactions that affect the trajectories
of the electrons. Further, it is also possible to study the 
novel two-particle
correlation, such as the Bell's inequality test~\cite{bell66}, 
using the entanglement proposed here. 

Finally, we add some remarks concerning the experimental feasibility
of our proposal, especially about the 
phase shift $\Delta\phi$ of Eq.~(\ref{eq:Delta-phi}) induced by
the Coulomb interaction.
In order to carry out the experiment, it is essential 
to get $\Delta\phi\sim\pi$. 
However, it is questionable if two spatially separated ballistic edge
channels suffer enough Coulomb interactions leading to $\Delta\phi\sim\pi$.
A possible solution to this question is to use two quantum Hall edge
states at a filling factor of 2. That is, with appropriate control of the
two edge states, one channel is used as an interferometer and the other
as a detector. Indeed, it has been shown that the Coulomb interaction
between the two edge channels at a filling factor of 2 is strong enough to
introduce a phase shift greater than $\pi$~\cite{neder05}.  
An alternative to get a strong Coulomb-interaction-induced phase shift
would be to insert a quantum dot in the interferometer as it has been
conducted in the earlier WP interferometer~\cite{buks98}.

In conclusion, we have proposed a possible realization of the
electronic quantum eraser by using a two-path interferometer and a 
quantum point contact detector in the quantum Hall regime. 
The Coulomb interaction between 
the interferometer and the detector induces a phase shift that enables
the which-path detection. We have shown that the which-path information
can be erased by a joint-detection measurement and then the hidden coherence
reappears. Furthermore, it is possible to choose the `particle-like' or
the `wave-like' behavior by appropriate control of the detector 
in a modified setup.  

\acknowledgements%
%We acknowledge helpful discussions with T.-G.~Noh and H.-S.~Sim.
%This work was supported by the ???? (Grant No. ??-??-???-???-???).

%%%%%%%%%%%%%%%%%%%%%%%%%%%%%%%%%%%%%%%%%%%%%%%%%%%%%%%%%%%
\begin{figure}
\centering%
\includegraphics*[width=70mm]{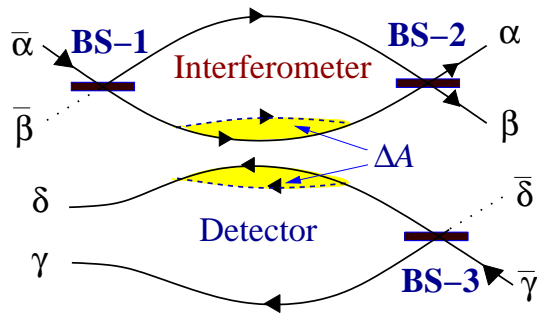}
\caption{A schematic diagram of the first setup for an electronic Mach-Zehnder
quantum eraser consisting of a Mach-Zehnder interferometer and a 
QPC detector. The Coulomb
interaction between the two electrons modifies their trajectories, giving
rise to a change of the areas, $\Delta A$. The 
interference of the output currents at leads $\alpha,\beta$ fully or partially
vanishes by the which-path information encoded in the phase shift 
$\Delta\phi$ of Eq.~(\ref{eq:Delta-phi}). The interference reappears by a 
joint-detection count (zero-frequency cross correlation) between a lead
of the interferometer ($\alpha,\beta$) and a lead in the detector 
($\gamma,\delta$).}
%\label{fig1}
\end{figure}
%%%%%%%%%%%%%%%%%%%%%%%%%%%%%%%%%%%%%%%%%%%%%%%%%%%%%%%%%%%
\begin{figure}
\centering%
\includegraphics*[width=70mm]{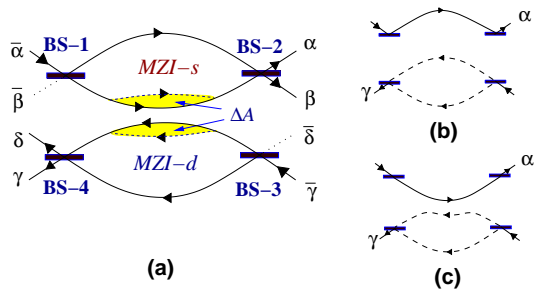}
\caption{(a) A schematic diagram of the 2st setup for an 
electronic Mach-Zehnder quantum eraser consisting of two coupled 
Mach-Zehnder interferometers with one being
used as an interferometer (MZI-$s$) and the other as a detector 
(MZI-$d$). In this geometry, the path information of the MZI-$s$ may be 
erased or restored by appropriate choices 
of the phase $\varphi_d$ in the detector. 
(b,c) Two indistinguishable processes for joint-detection at
leads $\alpha$ and $\gamma$.}
%\label{fig2}
\end{figure}
%%%%%%%%%%%%%%%%%%%%%%%%%%%%%%%%%%%%%%%%%%%%%%%%%%%%%%%%%%%

%%%%%% References

%%%%%%%%%%%%%%%%%%%%%%%%%%%%%%%%%%%%%%%%%%%%%%%%%%%%%%%%%%%

\begin{thebibliography}{99}

\bibitem{feynman65} R.~Feynman, R.~Leighton, and M.~Sands, {\em The Feynman
 Lectures on Physics} Vol.~III (Addison Wesley, Reading, 1965).

\bibitem{wooters79} W.~K.~Wootters and W.~H.~Zurek, \prd{\bf 19}, 473 (1979).

\bibitem{stern90} A.~Stern, Y.~Aharonov, and Y.~Imry,
 Phys.~Rev.~A~{\bf 41}, 3436 (1990).

\bibitem{scully91} M.~O.~Scully, B.-G.~Englert, and H.~Walther, 
 Nature~{\bf 351}, 111 (1991).

\bibitem{scully82} M.~O.~Scully and K.~Dr\"uhl, \pra{\bf 25}, 2208 (1982).

\bibitem{QE-exp} A.~G.~Zajonc, L.~Wang, X.~Ou and L.~Mandel, 
 Nature {\bf 353}, 507 (1991);  
 P.~G.~Kwiat, A.~M.~Steinber and R.~Y.~Chiao, \pra{\bf 45}, 7729 (1992);
 T.~J.~Herzog, P.~G.~Kwiat, H.~Weinfurter and A.~Zeilinger, \prl{\bf 75}, 3034
 (1995);
 T.-G.~Noh and C.~K.~Hong, J. Korean Phys. Soc.~{\bf 33}, 383 (1998); 
 Y.-H.~Kim, R.~Yu, S.~P.~Kulik, Y.~Shih and M.~O.~Scully, \prl{\bf 84}, 1 
 (2000).
% T.~Tsegaye and G.~Bj\"ork, \pra{\bf 62}, 032106 (2000).

\bibitem{buks98} E.~Buks, R.~Schuster, M.~Heiblum, D.~Mahalu, and
 V.~Umansky, Nature~{\bf 391}, 871 (1998).

\bibitem{sprinzak00} D.~Sprinzak, E.~Buks, M.~Heiblum, and
 H.~Shtrikman, Phys.~Rev.~Lett.~{\bf 84}, 5820 (2000).

\bibitem{kalish04} M.~Avinun-Kalish, M.~Heiblum, A.~Silva,
 D.~Mahalu, and V.~Umansky, Phys.~Rev.~Lett.~{\bf 92}, 156801 (2004).

\bibitem{aleiner97} I.~L.~Aleiner, N. S. Wingreen, and Y. Meir,
 Phys.~Rev.~Lett.~{\bf 79}, 3740 (1997).

\bibitem{gurvitz97} S.~A.~Gurvitz, Phys.~Rev.~B~{\bf 56}, 15215 (1997).

\bibitem{levinson97} Y.~ Levinson, Europhys.~Lett.~{\bf 39}
299 (1997).

\bibitem{hacken98} G.~Hackenbroich, B.~Rosenow, and
H.~A.~Weidenm\"uller, Phys.~Rev.~Lett.~{\bf 81}, 5896 (1998).

\bibitem{stodolsky99} L.~Stodolsky, Phys.~Lett.~B~{\bf 459}, 193
(1999).

\bibitem{buttiker00} M.~B\"uttiker and A.~M.~Martin,
Phys.~Rev.~B~{\bf 61}, 2737 (2000).

\bibitem{korotkov01} A.~N.~Korotkov and D.~V.~Averin, \prb{\bf 64}, 
 165310 (2001).

\bibitem{pilgram02} S.~Pilgram and M.~B\"uttiker, \prl{\bf 89}, 200401 (2002).

\bibitem{averin05} D.~V.~Averin and E.~V.~Sukhorukov, \prl{\bf 95}, 126803
 (2005).

\bibitem{kang05} K.~Kang, \prl{\bf 95}, 206808 (2005).

\bibitem{khym06} G.~L.~Khym, Y.~Lee, and K.~Kang, J.~Phys.~Soc.~Jpn.~{\bf 75},
 063707 (2006).

\bibitem{khym06a} G.~L.~Khym and K.~Kang, quant-ph/0606117.

\bibitem{hacken98a} G.~Hackenbroich, B.~Rosenow, and H.~A.~Weidenm\"uller,
 Europhys.~Lett.~{\bf 44}, 693 (1998).

\bibitem{optics84} See e.g., J.~R.~Meyer-Arendt, {\em Introduction to
 Classical and Modern Optics} (Prentice-Hall Inc., New Jersey, 1984), 2nd
 edition.

\bibitem{ji03} Y.~Ji, Y.~Chung, D.~Sprinzak, M.~Heiblum, D.~Mahalu, and 
 H.~Shtrikman, Nature {\bf 422}, 415 (2003).

\bibitem{buttiker90} M.~B\"uttiker, \prl{\bf 65}, 2901 (1990).

\bibitem{neder06} I.~Neder, M.~Heiblum, Y.~Levinson, D.~Mahalu, and V.~Umansky,
 \prl{\bf 96}, 016804 (2006).

\bibitem{neder05} I.~Neder, M.~Heiblum, Y.~Levinson, D.~Mahalu, and V.~Umansky,
 cond-mat/0508024 (2005).

\bibitem{englert96} B.-G.~Englert, \prl{\bf 77}, 2154 (1996).

\bibitem{note} Note that the distinguishability defined here is concerned with
 two particle processes while it is usually given in the context of the 
 single particle process as in \cite{englert96}.

\bibitem{bell66} J.~S.~Bell, Physics {\bf 1}, 195 (1964).
%%%%%%%%%%%%%%%%%%%%%%%%%%%

\end{thebibliography}
\end{document}